**Concerning two-gap ARPES data, the pseudogap, 2-*q* striping and magnetic circular dichroism from the negative-*U* HTSC perspective**

**John A Wilson**

H.H. Wills Physics Laboratory,
University of Bristol,
Bristol BS8 1TL.   U.K.

**Abstract**

Recent photoemission work on underdoped HTSC cuprates between $T_c$ and $T^*$, the pseudogap temperature, are interpreted within the framework of a previously developed two-subsystem, resonant negative-*U* scenario.  From this exposition, incorporating a developed 2**q** diagonal striping model, detailed understanding follows of the ARPES results, together with certain STM and far infrared optical data.  Identification of the origin of the magnetic circular dichroism and related effects exhibited by such materials below $T^*$ also appears to have been achieved.

**§1 Introduction and background to negative-*U* interpretation of cuprate HTSC behaviour**

In the 2006 Christmas issue of Science a couple of closely related ARPES papers were published by Tanaka *et al* [1] and Valla *et al* [2] of immediate relevance to the nature of the high temperature superconducting mechanism active in the hole-doped mixed-valent cuprates. Both papers probe the underdoped side of the phase diagram and the relationship borne there by the raised-temperature pseudogap condition to the low-temperature superconductively gapped state. In the underdoped regime these two kinds of DOS gapping, known to be similar in general angular form (both having nodes in the 45° direction, the basal Brillouin zone diagonals, and with maxima in the Fermi surface saddle/Cu-O bond directions) have been affirmed to be anticorrelated in magnitude. Upon reduction in hole count from optimal doping ($p \approx 0.15$), the specifically superconducting gap is seen to fall away, directly paralleling $T_c$, whilst, by contrast, the pseudogap grows monotonically. Such behaviour for the superconducting gap accords with data from specific heat [3], penetration depth [4], Andreev reflection [5], Raman spectroscopy [6], tunnelling [7] and other probes sensitive to the condensed state. What the new ARPES experiments provide is a clearer, more extended view of what was to be gathered from earlier ARPES [8,9] and energy-resolved STM work [10], in particular in regard to the strongly anisotropic distinctions current in the physics operative within the nodal and antinodal sectors. It has for some time been apparent that the action nearer the nodes is not especially unusual (except in degree) and that the remarkable input to the HTSC phenomenon arises with the special activity recorded in the antinodal, saddle-point regions. Nearer the nodes there exists rather classical fermionic behaviour, with a seemingly standard, though d-wave gapped Fermi surface, under impress of the superconductivity [11]. In the antinodal regions, however, there clearly is active a unique and chronic scattering which even in the normal state proves sufficient to precipitate disintegration of the Fermi surface there [12,13]. This developing incoherence imposes a positive Seebeck coefficient on all under- and optimally-doped HTSC systems below 300 K [14]. The Seebeck coefficient together with the Hall coefficient each exhibit highly anomalous temperature dependencies [15,16,17], and these partner very non-standard, highly anisotropic and strongly temperature dependent resistivity behaviour [18-21]. This abnormal quasiparticle behaviour centred upon the saddles has been discussed by the author in a whole series of papers since 1987 [22-33]. Those works expound cuprate HTSC phenomena in terms of inhomogeneous, negative-*U*, fermion-to-boson interchange. The latter is to be understood as arising within a system which is being driven to BCS/BEC crossover by 'accidentally' presenting a narrow (Feshbach) resonance between $E_F$ and the local negative-*U* state in question – specifically the double-loading fluctuation in electron content within a Cu$_{III}$ coordination unit from p$^5$d$^9$ to p$^6$d$^{10}$ (or in the notation of [23] $^8$Cu$_{III}$$^0$ to $^{10}$Cu$_{III}$$^{2\text{-}}$). In the most recent member of the above series of papers [33] the promoted superconducting condition around the nodes is as usual regarded as covered by B$_{1g}$ symmetry. At the same time the local pair action driving this low temperature state and centred upon the saddle hot spots is presented as holding geometrically related A$_{1g}$ symmetry, very suited to coordination unit breathing and closed-shell configuration physics. The bosonic local pair states established from the saddle region quasiparticles may either be held within the overall superconducting condensate or be external to it / excited from it. The two conditions for local pairs are perceived as engendering respectively the two distinct types of bosonic mode which have been sensed experimentally [see 31]. The first is in evidence in the mode-coupled physics registered in low temperature Raman [34,35], EELS [36], ARPES [37] and inelastic neutron scattering work [38,39], whilst the second, more $k$-dependant mode is met with in gigahertz spectroscopy [40] and energy-resolved scanning tunnelling microscopy [41,42]. The presence of local pairs to well above $T_c$ is supported now by the Nernst coefficient results [43,44]. The general positioning of the negative-$U$ state as being near to resonance with $E_F$ is, I have claimed in [28], extensively supported by data obtained from laser pump-probe experimentation [45,46] and other specialist optical techniques [47-49]. The extracted net value of the $Cu_{III}$-site Hubbard $U_{eff}$ is $\approx$ –3.0 eV per pair, or –1.5 eV per electron [28].

When underdoped, the negative-$U$ local-pair state ends up, it is adjudged, actually being stationed some tens of meV below $E_F$, at a binding energy represented here by cursive $U(p)$. With increased doping level, $p$, the local pair state is set slowly to rise (*i.e.* $U(p)$ will diminish) as the general level of metallicity increases, with the *local* Madelung potential around those copper centres transiently being driven towards formal trivalence (through the oxidative mixed-valent 'doping' in the 'double' Mott system) steadily being screened away, permitting the material to relax towards a more customary, homogeneous, Fermi liquid behaviour. In $(La_{2-x}Sr_x)CuO_2$ (LSCO) note, however, that the latter condition still is not fully attained even by $p$ = 0.3, beyond the point where HTSC itself has been relinquished [50]. Indeed in this range local moment magnetism starts to re-appear [51,52].

'Optimal doping' enfolds several factors. Firstly, because it occurs in all HTSC systems at virtually the same doping level ($p$ = 0.155), that fact has to reflect basic key geometrical constraints; namely i) in the random case, the first appearance of 2D percolation paths over those Cu-O coordination units immediately proximate to the charge centres in the random dopant array (see fig.4 in [23]), and ii) in the fully stripe-organized case, the crossing-point cluster-centring condition of fig.4 in [32] – see §3 below. Secondly one is confronted by the strong effect wrought by changes in counter-ion selection upon the size of the $T_c^{max}$ attainable in a particular system. What alteration in the counter ions incorporated (Ca, Sr, Ba, La, Hg, Tl, Pb, Bi, etc.) clearly involves is modification to the general level of ionicity and screening within the crucial $CuO_2$ layers. The way this proceeds is as follows. The ionicity of the counter-ion dictates the degree of covalent mixing of this ion's s- and p-states with the oxygen sublattice s- and particularly p-states, the principal contributors to the main valence band. These latter, predominantly oxygen-based states, then further hybridize with the copper d- and s-states by an amount controlled by the 'interim' energy separation. This hybridization then dictates the ultimate relative positions and widths of the states immediately relevant to HTSC. Primary among these are the $d_{x2-y2}$ symmetry, pd$\sigma$* (antibonding), basal-plane conduction-band states. The $d_{x2-y2}$ band in the present structures becomes totally elevated above the $d_{z2}$ symmetry, apical, antibonding $\sigma^*$ band by virtue of a Jahn-Teller distortion. This local distortion is effected under the complete filling of the $d_{z2}$ state with (sub)half-filling of its $d_{x2-y2}^*$ partner state in the $pd\sigma^*$ set of antibonding bands. The exhibited strong apical elongation of a CuO coordination unit not only indicates that locally the $d_{z2}$ state will be now significantly less antibondingly elevated than the basal $d_{x2-y2}$ state, but it implies too that the lattice has become basally compressed because the Jahn-Teller distortion is essentially volume conserving. The latter compression introduces a greater dispersion to the basal $d_{x2-y2}$ band than would be encountered in a less 2D-structured material. That increased dispersion brings in turn an enhanced tendency to resonant valence bond (RVB) formation within the conduction band and to a constraint upon free magnetic moments. Free moments of course constitute the principal agent for pair breaking in superconducting systems. From the above one is able to appreciate the delicate control which selection of the counter ions has over $T_c$. That choice dictates the absolute value of the chemical potential/work function and, via the level of screening attained at a given $p$ value, the degree of metallicity and magnetic moment suppression. At the present time $HgBa_2Ca_2Cu_3O_{16+\delta}$ with a certain amount of fluorine substitution provides the most advantageous combination towards elevating $T_c^{max}$ [53]. It is felt, however, that room still exists here for further ingenuity as the various parameters clearly are so sensitive and so strongly intertwined. The added basal compression secured by the insertion of additional fourfold coordinated $CuO_2$ layers plus Ca (in say HBCCO-1212 versus HBCO-1201) continues to be beneficial in raising $T_c^{max}$ only for just so long as such insertion does not lead to dispersal of the effect of the dopants away from the critical outer layers of the complete CuO sandwich [54, 55]. Note in the case of the Bi materials, $T_c^{max}$ for Bi-2201 stands far below that in Bi-2212 because bismuth imports particularly advanced covalent admixing. At the opposite, ionic extreme one finds ($Na_{2-x}Ca_x$)$CuO_2Cl_2$, where the metallicity has become so diminished that the Fermi surface begins seriously to disintegrate once below $x \sim 0.12$ [56], and magnetic [57] and checkerboard states [58,32] emerge to counter HTSC.

## §2. Regarding the recent ARPES results of Tanaka *et al* and Valla *et al.*

With this background we may now turn to examine the results and discussion appearing in the recent ARPES papers from Tanaka *et al* [1] and Valla *et al* [2] with the accompanying Comment by Millis [59]. From the position outlined above it is to be expected, and indeed found, that the situation upon moving into the underdoped region is controlled by two divergent energy scales, $\Delta(p)$ and $U(p)$. Under falling $p$, the instigating negative-$U$ state *binding* energy, $U$, (below $E_F$) of the antinodally created local pairs will increase, whilst $\Delta(\theta)$, the fermionic gapping induced around towards the nodal directions, will diminish, as the coupling between the negative-$U$ states and the fermionic open band falls away. Conversely towards optimal doping, $U$ and $\Delta$ (or rather $\Delta$ as extrapolated to the d-wave maximal value of $\Delta(\theta \to 0)$ within the saddles) draw together, as $U(p)$ becomes steadily reduced under metallic screening, while the strikingly small coherence lengths are expanded somewhat.

The peaks decorating the ARPES response have customarily been presented as classical coherence peaks, but such a view has been contested by the present author in [33]. There it is proposed that for these very highly correlated systems the ARPES emission process becomes coupled to the $c$-axis charge plasmon, strongly in evidence in FIR [60] and Raman [35] work. Notice how in the results of [1] the so-called 'coherence peak' remains at finite energy in the ARPES spectrum right round to the nodal point itself, and, furthermore, how it grows sharply in size and definition with rising $p$, as $T_c$ is advanced from 30 to 40 to 50 K in the UD $Bi_2Sr_2(Ca/Y)Cu_2O_8$ samples used. Once away from the nodal region and moving into the saddle region of advanced incoherence it is observed that this peak becomes much broadened. The peak's binding energy at the same time becomes so large (> 30 meV) that clearly it is not to be associated with $\Delta(\theta)$. If it were taken simply as $\Delta(\theta)$, the extrapolated saddle peak energies would correspond to $2\Delta^{max}/kT_c$ standing well in excess of what techniques more directly sensitive to the superconductivity itself support, namely $2\Delta^{max}/kT_c \sim$ 5½. Hence by the 'antinodal' direction the ARPES peak has lost any relation to $\Delta$, the gap feature of a classical, non-local pair (BCS) superconductor. It has come to lack due pair-coherent aspect, and in this it is mirroring the incoherence evident too in that region in the single particle spectrum. Both are expressions of the local pair creation and annihilation scattering current out of and back into these zone-boundary saddle-point states. As is then to be anticipated the broad peak feature present in the ARPES spectra at the saddles displays little or no temperature dependence across $T_c$, quite unlike with the peaking found once away from these key regions.

### §3. 2-q striping and localization.

The above incoherence, as noted earlier, is much more developed towards the more ionic limit of the HTSC systems, and in $(La_{2-x}Ba_x)CuO_4$ one meets with the well-known complete forfeiture of superconductivity close to $x = p = ⅛$. The latter event doubtless is encouraged by the structural complications exhibited in this particular system, with its HTT to LTO (at $T_{d1}$) to LTT (at $T_{d2}$) sequencing of low temperature, coordination-unit-tilted, superstructures [61]. The latter typify the ferroelectric/ferroelastic-type ground-state adjustments of ionic oxide systems, especially in those supporting Jahn-Teller distortion and orbital ordering. This is what brought Bednorz and Müller to examine these materials in the first place, with an eye towards soft-mode driven superconductivity. In consequence of being mixed-valent, there exists in the HTSC cuprates the added propensity to develop charge-ordering, present again in isostructural but Mott-insulating $(La_{2-x}Ba_x)NiO_4$. It is well known that in LBCO, etc. a residue of such 'stripe phase' behaviour still remains in evidence, dynamically at least, in particular in inelastic neutron and X-ray scattering results. What is found at $x = ⅛$ in LBCO (although not quite in LSCO) is that this charge ordering becomes a static one. The charge order brings in its wake magnetic ordering, taking appropriate $8a_o$ periodicity, as the two valence sites segregate out to leave areas of Cu(II) more strongly and systematically coupled. All this is well supported from extensive monitoring of the system by magnetic susceptibility [62], μSR [63], NMR/NQR [64], 2-magnon Raman work [65], etc., etc.. The elimination here at $p$ = ⅛ of RVB coupling, along with the removal of the low energy spin gaps which elsewhere characterize the HTSC cuprates, is able to be greatly furthered by perturbing the $CuO_2$ layers via Zn substitution [66] and also by the application of a $c$-axis oriented magnetic field [67].

Back in 1988 in [24] I queried whether what was happening in LBCO near $p$ = ⅛ might not actually be Wigner crystallization of the quasiparticles to create a 45°-rotated, ($2\sqrt{2}$ x $2\sqrt{2}$ $a_o$) charge order in empathy with the LTT lattice superstructure adopted below 60 K. However as the pseudogapping (at least in its lower energy reaches) has come to be seen as associated with the superconductive pairing, it has been questioned whether the incoherence developing in the Fermi liquid might not in fact be reflecting a crystallizing out of pairs rather than just quasiparticles [68]. In figures 1b,c the above two events are depicted against a background of charge striping such as would hold were the system at $p$ = ⅛ to behave in like manner to when away from this special, lattice commensurate, doping level. At the $x$ = ⅛ stoichiometry the overall $8a_o$ periodicity requires that sites internal to the domain wall stripes alternately carry Cu(II) and Cu(III) charge loadings, as in fig.1a. The stripe background to this figure immediately will be noted to have been presented as 2-**q** in form, rather than following the customary uniaxial 1-**q** dispensation. This is in keeping with an understanding of these events presented initially in [26,27] and subsequently much developed in [32]. In fig.1a one may immediately discern for these mixed-valent 'metals' the pertinent nodally-oriented 'rivers of charge'.

I am aware, after Fine's work along similar lines, lthough pursuing somewhat different arguments [69], that the matter of 2-**q** domaining recently has been re-addressed, in particular by Christensen *et al* [70]. From the latters' spin-polarized neutron diffraction data it is clear that 2-**q** domaining is indeed a viable option, although not one with basally collinear spin patterning such as was portrayed by Fine [69]. In my figures of [32] I chose not to engage, in the context of that paper, in addressing the matter of the anisotropic constraints operative upon the spin axes in the divalent domains. The spins were drawn there as if assuming the $c$-axis direction despite it long being known that in $La_2CuO_4$ itself they align almost in the basal plane [71]. For the mixed-valent striped case, as indicated earlier in figure 8 of [26], the probable outcome, gained in compliance with the magneto-elastic forces set up by the domain walls, is for the basically antiferromagnetic spin arrays either to be distorted hedgehog-like towards the walls or to be driven into circulations of spin directed as far as possible parallel to the walls. The existence of very strong magneto-elastic forces in LBCO in the vicinity of $p$ = ⅛ is well established from ultrasound data [72], and was discussed in §§4/5 of [26]. Such spin conformations now are displayed in fig.2 for the revised 45° orientation of 2-**q** domain walling advocated in [32]. It is my belief that these spin structures stand in agreement with the new refined neutron diffraction data secured by Christensen *et al* [70] – see §5 below.

Let us return then to the matter of the postulated Wigner crystallizations whether of quasi-particles or of pairs viewed against this setting of rotated 2-**q** domaining. To acquire uniform Wigner crystallization of the holes from the starting point of figure 1a calls simply for the relative phasing of charge alternation at the crossing points of the diagonal stripes to become slipped by π, with the 'surplus' hole thereby generated being then accommodated at a domain centre (fig.1b). It is, however, perhaps more in line with present work that what might crystallize are not the quasiparticles themselves, i.e. the holes, or even hole pairs, but, as was intimated above, *electron* pairs. The latter pairs, materializing at the Cu(III) negative-$U$ centres, potentially may become locked onto the stripe crossing-points and create thereby the 4x4 electron pair array shown in fig.1c. My feeling is that this in fact would constitute not so much a directly correlation-driven crystallization of pairs as one initiated by ferroelastic forces, paralleling in this the magneto-elastic forces that organize the spin arrays. As with the stripes themselves (and indeed with the very elevation of the $d_{x2-y2}$ σ* band to stand alone) the above forces emanate largely from the strong Jahn-Teller-based distinction between Cu-O coordination unit geometries under $d^8$, $d^9$ and $d^{10}$ site loadings. The tilts of the $^9Cu_{II}$ units in the LTO and LTT low temperature structures arise to accommodate there the strong apical (*c*-axis) elongation. The latter structural extension, whilst injecting such complexity now in LBCO at $x = ⅛$, becomes in $HgBa_2CuO_{4+δ}$ of great benefit, recall however, in contracting $a_0$ and so promoting RVB over ungapped, antiferromagnetic, spin coupling. In $p = ⅛$ LBCO, by contrast, the greater ionicity supports a resurgence of antiferromagnetism, which contributes much to preventing superconductivity being promoted over the Fermi sea.

It has to be emphasized here that the negative-$U$ scenario developed in [22-33] is one of electron pairs not hole pairs, the former associated with the attainment within a local negative-$U$ state coordination unit of the fully closed-shell $p^6d^{10}$ condition, the $^{10}Cu_{III}^{2-}$ state. The proximity to this closed-shell condition is what renders the cuprates unique (except for the rather similarly placed mixed-valent bismuthate family $(Ba/K)BiO_3$, etc.). It is this closed-shell physics (or, if you will, chemistry) which relates these systems to the phenomenon of charge disproportionation. However, as expressed from the start, lattice-constrained electron pairing emerging under static charge disproportionation is the last thing one in fact wants to encounter when hunting down HTSC. Static charge disproportionation in AgO, for example, becomes representable within the notation of [22] by

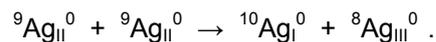

$$^9Ag_{II}^0 + {}^9Ag_{II}^0 → {}^{10}Ag_I^0 + {}^8Ag_{III}^0 .$$

Here the transfer across to the strongly semiconductive end-point is considerably furthered by virtue of the lattice structure adopted. In AgO, alongside the large monovalent coordination unit, one finds for the Ag(III) the often seen square-planar $d^8$ unit (as in semiconductive $d^8$ Pd(II)O, PdS, PdI_2, etc.). What a mixed-valent composition now accomplishes for the HTSC cuprates and the bismuthates, in consequence of their intrinsic structural disorder, is a forestalling of any such drive to take on a statically disproportionated lattice structure. As the amount of mixed-valent doping with its attendant metallicity is diminished, any latent tendency towards displaying charge disproportionation will become less frustrated. What appears in fig.1c for $p = ⅛$ LBCO could in fact be viewed as a frozen disproportionation wave rather than an electrostatically driven quasiparticle pair crystallization. By concentrating upon a 4x4 electron-pairing square repeat of 16 unit cells, it is readily seen that the process indicated amounts to adjusting from $(2d^8 + 14d^9)$ to $(3d^8 + 12d^9 + 1d^{10})$; i.e. a net shift from $2d^9$ to $(d^8 + d^{10})$ – very much as under disproportionation.

Both above pictures indeed then hold undesirable characteristics. While a loss of superconductivity could well ensue from such charge 'crystallization', it would come at the price of disassociating the ⅛ material from its flanking compositions. In particular there would ensue an overriding of the continuity evident in the IC neutron *spin* diffraction results, below as above $T_{d1,2}$, and, what is more, there is an implication that the outcome should be far less metallic than actually proves the case. It is felt that the above notions of Wigner crystallization have emerged from over-emphasis upon the holes and upon the stripe array. If we refocus on the interior of the domains, and on the electrons based there, we will arrive at a much more satisfactory perspective on events.

## §4. Consequences for the optical, magnetic, transport and tunnelling results.

What the resurgence of magnetic behaviour at the congregating $d^9$ sites near $x = ⅛$ signals is a move towards localization for the electrons and specifically those of the domain. While $\rho(T)$, $R_H(T)$ and $S(T)$ do not indicate the onset of full-blown Mott localization, it is very evident that a form of weak localization is being encountered. While $\rho(T)$ shows just a small upward step discontinuity of <5% as $T$ passes below $T_{d2}$ [73], by contrast $R_H(T)$ falls more sharply down through zero than is found away from $p = ⅛$ [74], and $S(T)$ [75] also advances its departure from mounting positive values. What is implicated is that the inner domain is losing coherent electronic contact with the stripes, as the lattice strain at the domain boundary grows under the developing charge segregation. The Cu(II) patch is coming to define a self-trapped, meso-scale, charge defect, with a binding energy of the order of a few $kT$. (Note $T_d$ ≈ 60 K is ≡ 5 meV). One might ask here if there is any evidence for such self-trapping showing up in the far-infrared (FIR) optical spectrum – and the answer is yes. The single crystal reflectivity work on $x = ⅛$ LBCO (vs LSCO) recently published by Homes *et al* [76] would lend full support to such an interpretation. To fit their FIR data calls for the optical conductivity to be apportioned between a coherent, Drude-like component, such as would encompass the near-nodal charge motion, and a significant incoherent constant component, appropriate to deal with the near-axially-directed **k** states. A Kramers-Kronig analysis made to the above format uncovers a very significant spectral weight loss from energies below 200 $cm^{-1}$ (≡ 25 meV or 300 K), this setting in sharply at $T_{d2}$. $n_D$, the coherent charge density, by 4 K becomes diminished to just 25% of its value at 60 K. At the same time, as very evident from the rapidly sharpening very low frequency Drude tail to the optical conductivity, the optical scattering rate extracted for these coherent states accelerates its general reduction once $T < T_{d2}$.  At $T_{d2}$ note that $1/\tau_D$ (equal there to 108 cm$^{-1}$) is equivalent to $2\frac{1}{2}kT_d$.  Such a scattering rate translates into a characteristic time of around $2\times10^{-12}$s, and speaks of a process that is phonon limited.  As the rate-determining charge exchange between the domains and stripes starts to freeze out strongly below $T_{d2}$, that change becomes manifest in the accelerated *reduction* in scattering rate of the remaining coherent electrons.  The latter nodally-directed stripe carriers now are less perturbed by the ingress of heavy electrons from the domains.  Of course, as supply of such electrons to the negative-$U$ centres dries up, so too will the rate of generation of local pairs, and with this terminates the potential to project superconductivity back into the domains.  The HTSC cycle collapses.

What are the observed consequences of the above condition for the non-superconducting $x = \frac{1}{8}$ material?  It has become clear that the events seen in LBCO are not simply the outcome of the adoption of LTT superlattice structuring alone.  Indeed the effects plainly extend through to temperatures appreciably above $T_{d2}$.  Correspondingly in LSCO, while the application of a magnetic field is known to have virtually no effect upon the lattice softening associated with the incipient low temperature superlatticing [72], it has been observed to have a big effect upon the magnetic and superconductive behaviour around $x = \frac{1}{8}$ [67].  The LTO and LTT crystal structure transformations per se clearly are little affecting the band structure at $E_F$, as is evidenced by the $\rho(T)$, $S(T)$ and $R_H(T)$ plots displaying only rather subtle changes at $T_d$.  Indeed the $x = \frac{1}{8}$ transport data step not greatly out of line in general magnitude with what is found for the flanking compositions at which superconductivity arises.  What do alter are the EXAFS-recorded Cu-O bond lengths, as is to be anticipated for a charge localization process quasi-static in nature [77].  The main effect of the low temperature changes around $x = \frac{1}{8}$ is, one gathers, to introduce a somewhat augmented localization, and with this there arrives an increase of singlet pair breaking for the Fermi sea based condensate.  It is not that the negative-$U$ pair states have disappeared or have been markedly shifted in energy away from resonance, and accordingly rendered ineffective at inducing superconductivity.  It becomes apparent that the $\frac{1}{8}$ condition principally entails a slower inter-subsystem charge fluctuation rate plus a faster depairing rate.

Whilst the $B_{1g}$ d-wave superconductive gap becomes suppressed, that does not mean the $A_{1g}$ gapping associated with the negative-$U$ state will likewise have vanished.  Long ago it was revealed by Maggio-Aprile and coworkers [78], employing STM to examine the situation inside magnetic field vortices in HTSC materials, that, with the re-institution therein of magnetic behaviour, the quenched superconductivity and associated loss of gapping are not accompanied by any simultaneous elimination within the vortices of the larger energy 'pseudogap' peak.  The latter negative-$U$ feature is witnessed to remain in evidence there at the same energy as is observed in the field-free condition.  It is this same 'pseudogap' feature that now is reported on by Valla *et al* [2] in the new ARPES work for $x = \frac{1}{8}$ LBCO.  The observed peak energy there of 20 meV stands entirely in agreement with the 'pseudogap' energy from the antinodal ARPES spectrum of underdoped LSCO recorded by Tanaka *et al* in [1], and in evidence there to well above $T_c$.  It is a feature which in the underdoped material stands well above the $2\Delta$ values specifically relating to the onset of superconductivity itself. It appears to the author that there is no problem whatsoever then in embracing all these findings within the scope of the present, inhomogeneous, negative-$U$ scenario.

**§5. Encompassing the magnetic circular dichroism and related results.**

A further outcome seeming to emerge from the above considerations is an explanation for the magnetic circular dichroism (MCD) exhibited by underdoped HTSC cuprates [79]. The motivation for the MCD experiment came from Varma [80] upon contemplating whether the pseudogap state might involve a subtle symmetry breaking, with associated order parameter, upon engagement at $T^*$, rather than it simply being a fluctuational condition. Because the lattice structure/symmetry has long been deemed as unchanging at $T^*$, the natural symmetry breakage postulated was of time reversal symmetry: TRSB is customarily associated with spin. Since $T^*$ would appear from neutron diffraction not to be a standard magnetic transition, it was proposed that, if the experiment were to detect a weak symmetry breakage, it may actually be indicative of spontaneous current-loops occurring at the unit cell level. The latter would, of course, be associated with weak local magnetic moments, and probably also with zero net magnetization, accounting thereby for the absence of any obvious betrayal hitherto of their presence. The fact that Kaminski *et al* in [79] do appear in their circularly polarized ARPES experiment on underdoped BSCCO to have detected a small but well-structured MCD response (both below $T^*$ as $T_c$), which is absent for overdoped BSCCO, has provided some support for there being an exotic current-loop precursor condition to UD HTSC. However, it is my belief now that interpretation of this experimental finding is a good deal more prosaic, and that it is associated with the spin structures which develop in conjunction with the diagonal 2-**q** charge striping met with above. Moreover the MCD results would, it looks, permit us to make selection between the two magneto-elastically constrained possibilities presented in figure 2, this in favour of fig.2b – the circulatory spin pattern.

An additional result to incorporate prior to considering the details of the above MCD work is that evidence for relatively weak magnetic ordering activity now has been secured below $T^*$ in underdoped YBCO too. The spin diffraction recently registered there by Fauqué *et al* [81] via elastic spin-polarized neutron scattering is quasi-static on the neutron diffraction time-scale of $10^{-12}$ s. $T^*$ appears thus to be where stripe ordering begins to become organized. The 'ordered' moments reported up now in this region are only around 0.1 $\mu_B$, but there are manifest quite appreciable magnetic coherence lengths of 50 Å plus (i.e. three to four diagonal domains). As noted already, the application of a magnetic field enhances site moments, and it is accordingly not surprising that Shi *et al* [82] from their measurement in a field of 8 tesla of the infra-red Hall effect make additional report of Faraday rotation and circular dichroism in underdoped LSCO right through to 300 K. The new neutron work by Fauqué *et al* [81] would indicate that in YBCO the freezing magnetic moments tilt very substantially out of the basal plane, and in fact probably lie closer to the $c$-axis than to the $a,b$ plane. YBCO with its chains is, of course, not the archetypal HTSC material, but clearly one can contemplate the spins being strongly tilted out of the basal plane too in the other HTSC systems, under the strong magneto-elastic action of the stripes. Hence figures 2a,b show only the orientation of the horizontal components of the site spins. These arrays settle steadily towards a 'spin-gapped' ground state.

We now may examine in detail what Kaminski *et al* observe in the MCD ARPES experiment [79]. Their set up was so configured that across a standard *c*-axis structural mirror plane the relative intensities of right- and left-circularly polarized (RCP/LCP) ARPES signals will swap over at **k**-vectors lying precisely on the mirror plane. *Over*doped samples indeed yield just such a 'null' result, both for the axial and the diagonal mirror planes of the (pseudo-tetragonal) crystal structure. However the various *under*doped samples examined all generate, once below $T^*$, an MCD signal indicative of TRSB. This symmetry breakage is deduced to relate uniquely to the axial (saddle direction) mirror of the basic structure, and not to the diagonal mirror. The distinguishing MCD response involves a small transverse offset (2.3°) to the RCP/LCP equal intensity crossover **k**-point away from the vertical mirror plane. Right at the mirror there is then present an intensity difference between RCP and LCP ARPES signals, and rotating the sample by $^\pi/_2$ sees the sign of this intensity difference reversed. These results show that the magnetic constraints on the signal are such that TRS breakage of the basic symmetry arises with regard to the axial mirror of the crystal structure but not to the diagonal mirror. Inspection of figures 2a and 2b reveals that it is the latter of these two circumstances, the circulatory arrangement, which must then hold. In retrospect this is not an unreasonable outcome. We observe with choice 2b that the spins indeed reflect across the diagonal plane but not across the axial Cu-O-bond/saddle mirror, just as the ARPES experiment would convey. With 2a the converse holds. Note that the spin arrangement in 2b (as in 2a) manifests antiphase (spin discommensurate) character along the *axial* directions across the stripe crossing points, as deduced in the early neutron diffraction work. Across individual diagonal stripes there exists a mirror relationship evident in fig 2b that is absent in fig.2a. The repeat pattern of spin domains is in both cases face-centred and bears a $\sqrt{2}$ relationship to the *hole charge* patterning. The latter is face-centred itself, this time with regard to the overall $8a_o$ axial supercell. It is this face centring that is responsible for the systematic absences witnessed in both spin and charge diffraction spotting, as pointed to in [32].

## §6. The way forward.

It is fascinating to see again how the twelve inherent aspects to HTSC cuprate physics pointed to in 1987 in [22/23] in practice work through to build a situation of such rich complexity. From this platform it is surely time now to probe once more the actual dynamics of the superconducting process - the heart of the matter – via the employment of pump-probe laser techniques. The earlier work of [46] and [48] (and what was made of these in [28]) has been added to recently by the results of [83], and this work should now be made to embrace what is known regarding the stripe condition in optimally and underdoped material. The key aim must be to clarify the source of the observed, three-component, dynamical behaviour, and to ascertain whether its relationship to the inhomogeneous two-subsystem/domain structuring of these materials indeed can continue to sustain the conclusions drawn in [28], upon which so much in the current paper and its predecessors rests. It is to be hoped finally that somebody will take up the challenge made in [33] to examine theoretically the present negative-$U$ stand-point by means of cluster dynamical mean field techniques.

**Acknowledgments.**

The author gratefully acknowledges receipt of a University Fellowship, permitting this work to be carried forward. My thanks go again to the members of the low temperature group in Bristol for their shared enthusiasm over all matters relating to HTSC.

**Postscript** (27.06.07)

Since the original version of this manuscript was released as *arXiv:cond-mat*/0703251 a couple of papers have been published relating directly to §5 and which are most supportive of the picture presented there. Firstly the analysis S. Di Matteo and M.R. Norman (2007 *arXiv:cond-mat*/0704.0599) give of the X-ray circular dichroism earlier reported for the K-edge spectra of BSCCO would endorse that the spin structuring has to be of a more complex type, such as is developed in §5. Secondly the new paper from B.V. Fine (2007 *Phys. Rev.* B **87** 060504(R)), using direct diffraction analysis, upholds the conclusion reached above that the spin patterning of figure 2b indeed is able to encompass the neutron diffraction data coming from the spin-polarized experiments on $(La_{1.48}Nd_{0.40}Sr_{0.12})CuO_4$ by Christensen *et al* [70]. Attention finally is drawn to the latest works by S.K. Adhikari *et al* (2007 *Physica* C **453** 37) and T. Stauber & J. Ranninger (*arXiv*:0706.3143[cond-mat.supr-cond]) on theoretical negative-$U$/crossover modelling with regard to the HTSC phenomenon.

**Figure captions**

**Figure 1**

(a) The 2-**q** diagonal stripe array for $p = \frac{1}{8}$, as introduced in [32]. The $8a_o$ supercell detected for LBCO by elastic X-ray and neutron diffraction sets the scale to this figure, in which the copper atom sites predominantly feature. Cu(II) sites in the domains between the stripes are marked **+** and **o** to indicate their antiferromagnetically coupled status, whilst in the stripes non-magnetic Cu(II) sites alternate with Cu(III) 'hole' sites. The latter sites are framed by small squares marking their four basal coordinating oxygen locations. The stripe crossing points are each decorated by four holes to yield an overall supercell that is face-centred. The inner domains are alternately dominated by 'up' and by 'down' spin electrons, as indicated by the areal shading, once again defining a face-centred array (but 45°-rotated). The $\sqrt{5}a_o$ circle at the bottom right marks the centre of negative-$U$ activity at the stripe crossing-points with their enhanced Madelung potential. The stripes provide the diagonal (nodal) coherent 'rivers of charge', whilst the domains supply heavy/incoherent $x$- and $y$-axis saddle electrons to the negative-$U$ centres for pairing. This patterning becomes completely frozen in $x = \frac{1}{8}$ LBCO below the LTT tilt transition at $T_{d2}$, resulting in the loss of HTSC behaviour.

(b) Postulated Wigner crystallization of quasiparticle holes at $p = \frac{1}{8}$, as set against background of the diagonal 2**q**-striping of figure 1a. To reach this pattern from figure 1a requires that the Cu(II)/Cu(III) phasing between stripes be slipped by $\pi$ so that a hole sits at the stripe crossing points. The 'surplus' hole per $8a_o$ cell then is accommodated at the supercell centre to reach the regularized hole array presented.

(c) Postulated Wigner crystallization of *electron* pairs at $x = \frac{1}{8}$, again employing 2**q**-diagonal striping as reference background. This condition would require the net transfer of two electrons into those Cu(III) sites in figure 1b formerly occupying the symmetric sites at the domain corners and centres as is indicated. Besides being symmetry-wise rather odd, this transfer effectively amounts to a disproportionation process for which there is no evidence elsewhere in copper chemistry (unlike silver and gold).

**Figure 2**

(a) Spin array of figure 1a as constrained by Jahn-Teller supported magneto-and ferro-elastic forces established at boundaries of mixed-valent diagonal striping. The small arrows show the directed horizontal components of the spins at the domain sites. In this hedgehog array of spins the latter are constrained as far as possible to set orthogonal to the stripes. Alternate domains, following the experimentally endorsed spin discommensuration sequencing of figure 1a, are dominated by outwardly and then inwardly directed basal spins. Polarized neutron experiments on YBCO [82] suggest that the spins actually are aligned closer to the vertical than to the horizontal. The saddle mirror remains intact with this spin array, but the original diagonal lattice mirror is broken by this pattern.

(b) Partner figure to 2a with the spins now constrained as far as possible to set with their basal components parallel to the charge striping. These circulations of spin alignment this time conserve the diagonal mirrors but break the axial mirror operations. It is this circumstance which, unlike that of fig.2a, matches the magnetic circular dichroism data obtained by Kaminski *et al* [79].

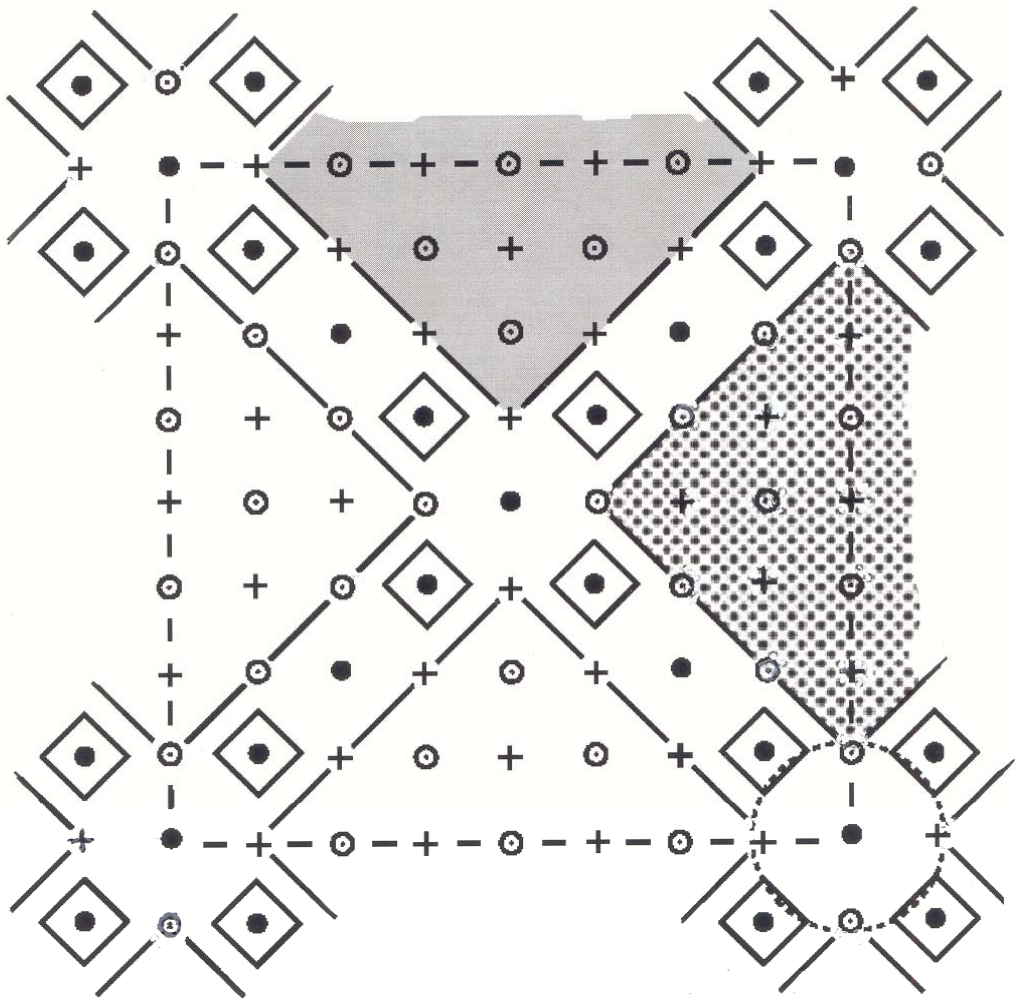

f1a

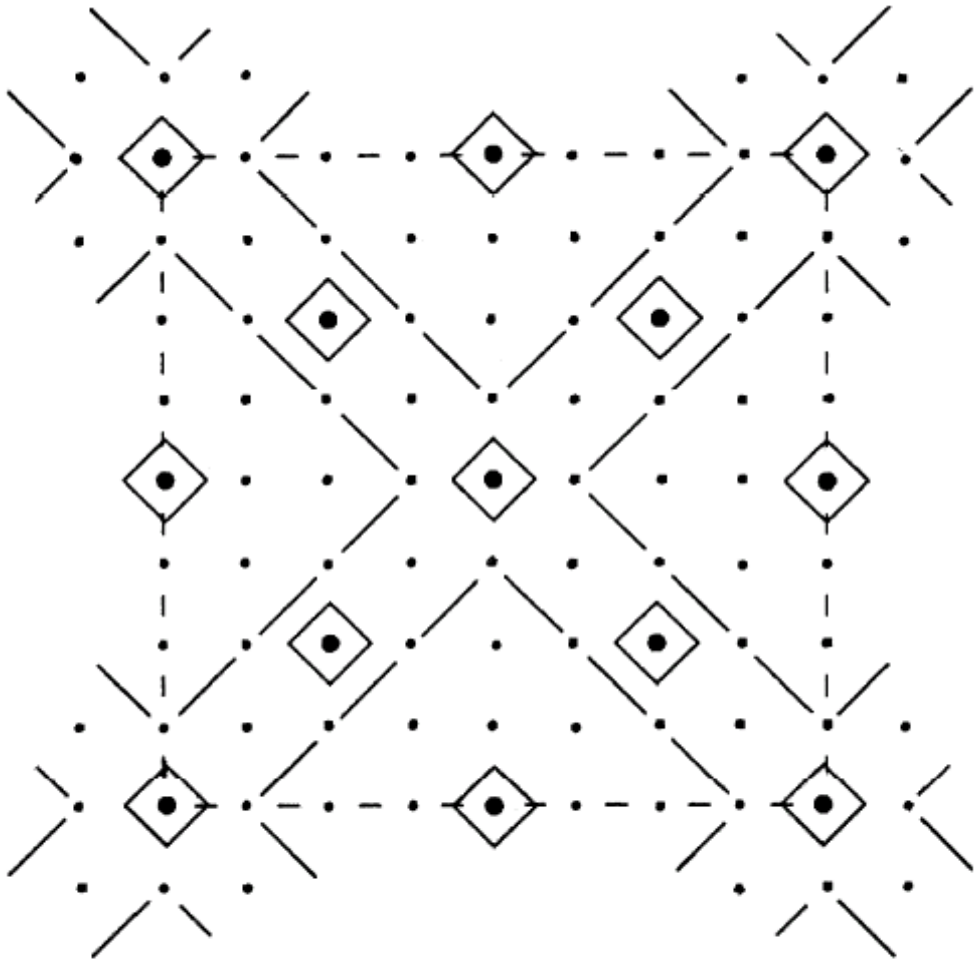

f1b

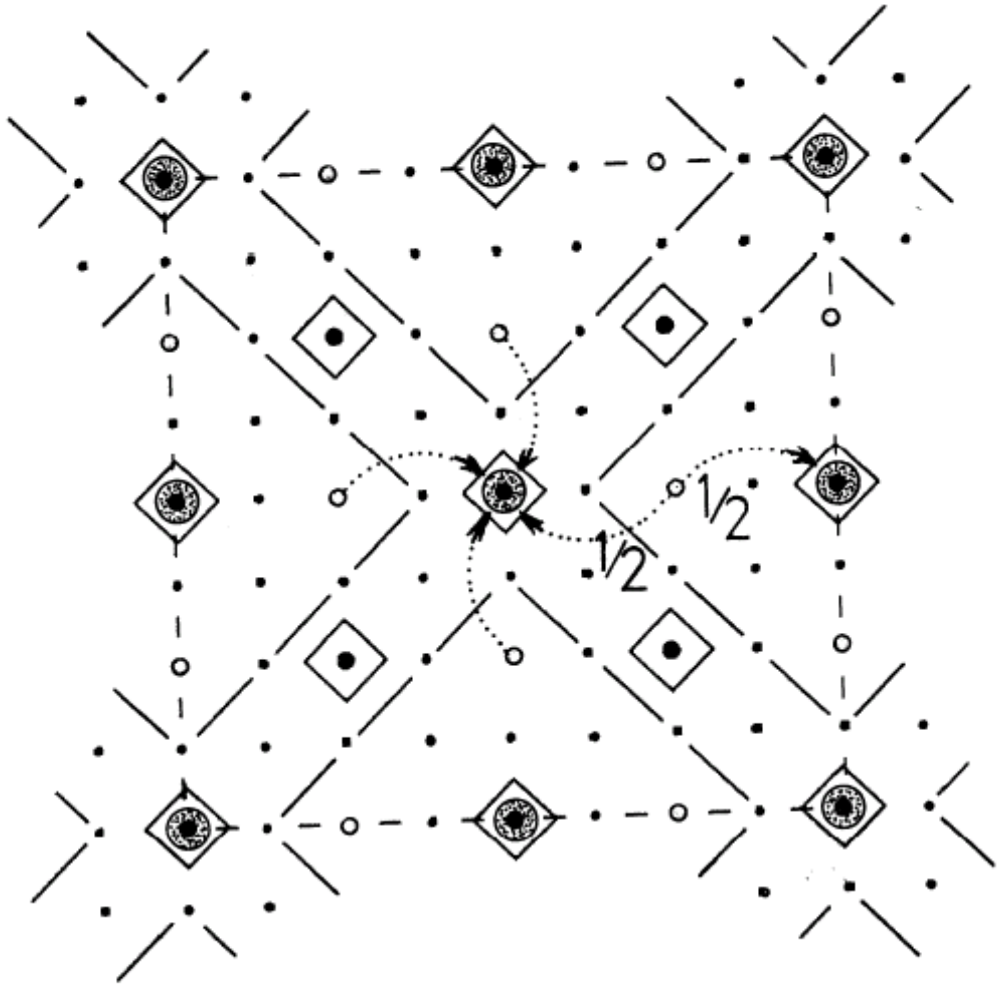

f1c

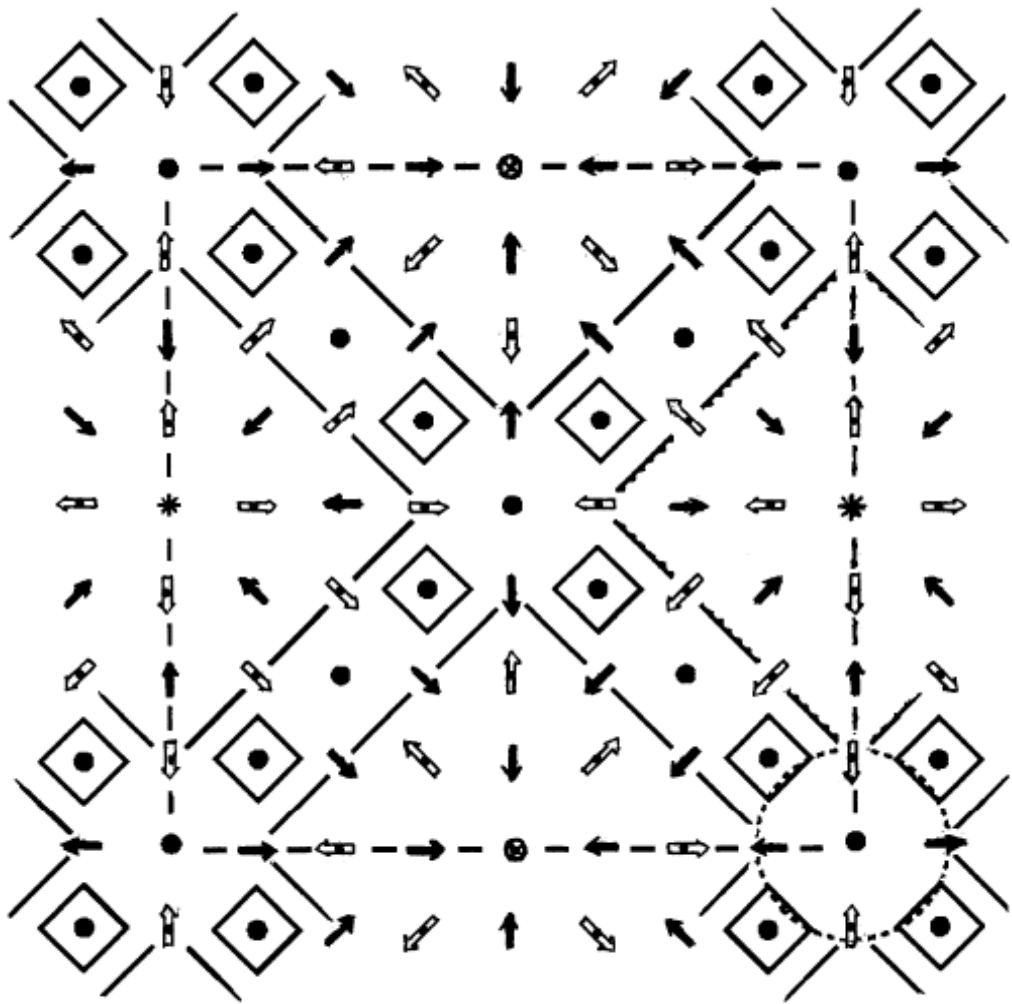

f 2a

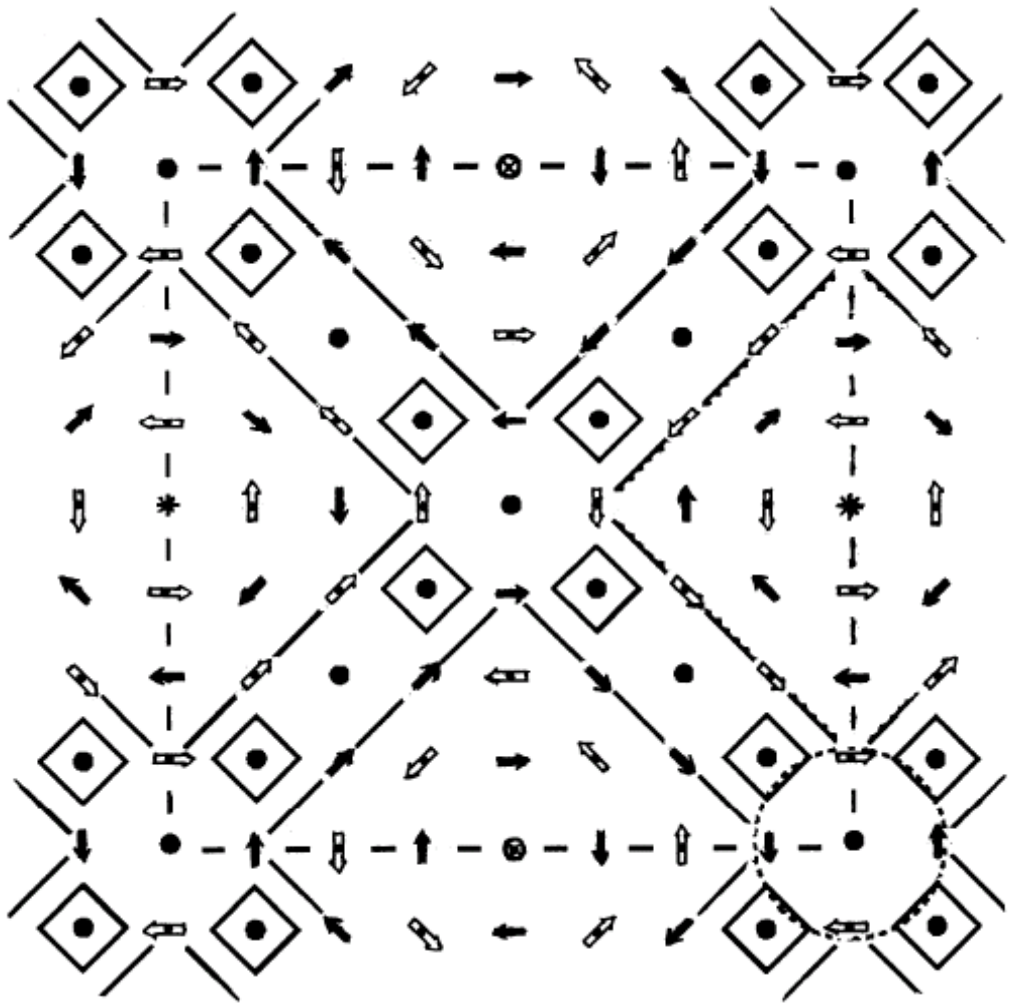

f2b